\documentstyle[11pt,epsf]{article}
\catcode `\@=11
\def\be{\begin{equation}}
\def\ee{\end{equation}}
\def\bd{\begin{displaymath}}
\def\ed{\end{displaymath}}
\newcommand{\ba}{\begin{eqnarray}}
\newcommand{\ea}{\end {eqnarray}}
\newcommand{\nn}{\nonumber}
\newcommand{\ket}[1]{|{#1}\rangle}

\def\a{\alpha}
\def\b{\beta}

\def\d{\delta}

\def\pr{\prime}
\def\ra{\rightarrow}

\def\inf{\infty}

\def\half{\frac{1}{2}}

\begin{document}
\thispagestyle{empty}
\pagestyle{plain}
\begin{center}
\large
{\bf A novel exponent in the Equilibrium Shape of Crystals}
\end{center}
\begin{center}
\normalsize
Silvio Renato Dahmen \footnote{Sloane Physics Laboratory, Yale
                   University, New Haven, CT 06520, USA},
Birgit Wehefritz \footnote{Physikalisches Institut, Universit\"{a}t
                 Bonn, Nu\ss allee 12, D-53115 Bonn, Germany}
and Giuseppe Albertini \footnote{Physics Dept. of Milan University,
                   via Celoria 16, 20133 Milano, Italy}
\\[10mm]
{\bf Abstract}
\end{center}
\small
\noindent
%
%
A new exponent characterizing the rounding
of crystal facets is found by mapping a crystal surface onto the
asymmetric six--vertex model (i.e. with external fields $h$ and $v$)
and using the Bethe Ansatz to obtain appropriate expansions of the
free energy close to criticality. Leading
order exponents in $\delta h$, $\delta v$ are determined {\it along the
whole phase boundary and in an arbitrary direction}. A possible
experimental verification of this result is discussed.
%
%
\rule{5cm}{0.2mm}
\begin{flushleft}
\parbox[t]{3.5cm}{\bf PACS numbers:}
\parbox[t]{12.5cm}{05.70.Jk, 64.60.-i, 64.60.Fr, 75.10.Jm}
\parbox[t]{3.5cm}{\bf Keywords:}
\parbox[t]{12cm}{equilibrium crystal shapes, six--vertex model,
critical exponents, Bethe Ansatz}
\vskip 1.0cm
Yale Preprint Nr. YCTP--P24--97\\
Bonn Preprint Nr. BONN-TH-98-05
\end{flushleft}
\small

\newpage
\pagestyle{plain}
\setcounter{page}{1}
Along a first--order coexistence line the shape of a solid inclusion
in equilibrium with its fluid is determined by the Gibbs--Curie
constrained minimization condition: for a fixed
volume $V$ of the inclusion, the equilibrium surface $\Omega =z(x,y)$
is such as to minimize the surface free energy
\be
F(T) = \int_{\Omega}\tau (\vec{n},T)\; d\Omega = {\mbox{minimum}} 
\label{eq1}
\ee
where the surface tension $\tau$ depends on the orientation of $\Omega$
$\vec{n} =\vec{\nabla} z(x,y)$ and temperature $T$. To solve the
constrained variational problem to determine $z(x,y)$ one
introduces a Lagrange multiplier $2\nu$
\bd
\delta\biggl\{ \int [ f(p,q) -2\nu z(x,y)] dx dy\biggr\}=0
\ed
$p=\frac{\partial z}{\partial x}$, $q=\frac{\partial z}{\partial y}$
are the slopes of the surface and $f(p,q,T) = \tau (p,q,T)
\sqrt{1+p^2 +q^2}$ is the specific surface free energy projected onto the
$(x,y)$--plane. The solution reads \cite{landau1,andreev}
\be
\nu z = \overline{f}(-\nu x, -\nu y)
\label{eq4}
\ee
where the Legendre transformed potential
$\overline{f}$ is defined through \cite{andreev}
\be
\overline{f}(\eta_x ,\eta_y ) = f(p,q) -p\eta_x - q\eta_y
\label{eq3}
\ee
with $\eta_x = \frac{\partial f}{\partial p}$,
$\eta_y = \frac{\partial f}{\partial q}$ the
surface tilting fields conjugate to the slope variables $p$
and $q$. 
Thus the determination of $\overline{f}$ yields the
equilibrium surface $z(x,y)$ directly, up to some
overall irrelevant constant. 
The main question however is how much of the physics
can be captured by this description. Crystals are believed to be
entirely faceted at $T=0$. As the temperature is raised, thermal
fluctuations bring about a rounding of the edges where
planes with different indices $(hkl)$ meet. At the
roughening transition temperature $T_{\mbox{\tiny R}}$ the remaining
planar portion of $(hkl)$ disappears entirely and the face becomes
rough. Different facets may have different $T_{\mbox{\tiny R}}$'s.
It is clear that questions concerning
the exact value of $T_{\mbox{\tiny R}}$ or exponents characterizing
the macroscopic smoothing out of the edges cannot be addressed in
a satisfactory way unless one has a microscopic model from which
$\tau$ can be derived from first principles.
This problem has been tackled with success by several authors
mainly by using a connection between the theory of equilibrium shapes and
two--dimensional lattice models, with the asymmetric
six--vertex model playing a major role
\cite{van_beijeren,rottman_wortis,jaya_saam,nolden}.
This system of interacting
dipoles on a lattice is the natural extension of the symmetric problem
through the inclusion of external fields $h$ and $v$
\cite{lieb_wu}. It is a classical paradigm of models
soluble through the Bethe Ansatz and an exact expression for its
free energy has been known for many years \cite{syy}.
Originally introduced in the context of equilibrium crystal shape
problems (ECS) to
describe the $(001)$ facet of a bcc crystal \cite{van_beijeren}, it
was later extended to cover other facets on different lattices as
well. In this way very comprehensive results have been obtained
regarding the form of the equilibrium surface as well as its thermal
evolution \cite{jaya_saam}.
In particular there is a region in the $(h,v)$ plane bounded by a
curve $\Gamma$ and containing the origin $(h=0,v=0)$, which defines the loci
of points where the free--energy retains its zero--field value \cite{syy}.
This implies that $z$ is a constant over an
entire region of the parameter space. Beyond $\Gamma$ the free energy
changes continuously as a function of the fields.
Appropriate expansions of the surface free energy around
$\Gamma$ show a leading order exponent for the scaling of
$z(x,y)$ as one approaches the boundary of the $(110)$ fcc facet along
the $y=0$ direction \cite{jaya_saam}
\be
z \sim (x-x_c )^{\frac{3}{2}} + {\cal O} (x-x_c )^{2}
\label{expoente}
\ee
This result has been confirmed experimentally
\cite{rottman_wortis2, carmi}.
However, even though a great
deal has been learned from exact analytical methods, their handling is no
trivial matter and previous results were somewhat limited in their scope by
rather involved technical difficulties. In spite of the fact that it had
already been conjectured that (\ref{expoente}) should hold on {\it the entire
facet boundary} a proof of this result still eluded us \cite{jaya_saam}.
In this letter we give such a proof based on the expansion of
the free energy of the asymmetric six--vertex model via the Bethe Ansatz,
introduced in \cite{silvio2,giuseppe} as a generalization of
previous methods \cite{lieb_wu,nolden}. The main results can be summarized
as follows: the exponent
$\frac{3}{2}$ dominates the rounding in all directions
{\it but} the tangential one, in which case a new exponent $3$ dominates.
These results hold for the whole boundary.

For the sake of clarity, we consider one particular
geometry, namely
that of the $(110)$ facet of an {\it fcc} lattice \cite{geometry}.
One may associate an energy to the
surface arising from the cleavage of an infinite ideal
crystal through the plane $(hkl)$ based solely on the number and type
of bonds broken per primitive cell. This is the simplest model possible
and yet one which allows an exact handling of the equations that follow.
Consider an fcc lattice (fig. $1$) with energies $-J_1$
and $-J_2$ assigned to nearest-- and next--to--nearest neighbor
bonds respectively ($J_1>J_2 >0$). The $T=0$  surface tension
reads \cite{mckenzie}
\be
\tau (\vec{n}_{hkl}) =\frac{\vec{n}_{hkl}}{V^{pc}|\vec{n}_{hkl}|}\cdot
\biggl( J_{1}[210] + J_{2} [111]\biggr)
\label{eq6}
\ee
where $\vec{n}_{hkl} = h\widehat{x} + j\widehat{y} +k\widehat{z}$ is the
vector normal to the plane $(hkl)$ and $V^{pc}$ is the volume of the primitive
fcc cell. Voids (excitations in the bulk)
as well as overhangs (height differences between neighboring columns of
atoms bigger than some unit length) are strictly forbidden
(solid--on--solid condition). The connection between the $(110)$ facet
and the vertex model is shown in figure $1$. From
(\ref{eq6}) one may associate the following energies to the vertices
\cite{jaya_saam}
\ba
e_1 &=& J_2 +\frac{A}\eta_x = J_2 - H - V\nonumber\\
e_2 &=& J_2 -\frac{A}\eta_x = J_2 + H + V\nonumber\\
e_3 &=& J_1 -\frac{A}{\sqrt{2}}\eta_y = J_1 - H + V\nonumber\\
e_4 &=& J_1 +\frac{A}{\sqrt{2}}\eta_y = J_1 + H - V\nonumber\\
e_5 &=& e_6 \; =\; 0 
\label{eq7}
\ea
Here $A=\frac{a^2}{\sqrt{2}}$ is the area of a vertex
(for a lattice constant $a$) and
\ba
H &=& -\frac{A}{2}\biggl(\eta_x -\frac{\eta_y}{\sqrt{2}}\biggr)
\nonumber\\
V &=& -\frac{A}{2}\biggl(\eta_x +\frac{\eta_y}{\sqrt{2}}\biggr)
\label{eq8}
\ea
Note that this parametrization is slightly different than
the one used in \cite{jaya_saam} but is consistent with figure 1.

The free energy (\ref{eq3}) follows straightforwardly \cite{jaya_saam}
\be
\overline{f}(\vec{\eta},T) = f(110) + f_{6-vertex}(J_1,J_2,H,V,T)
\label{eq8a}
\ee
A configuration of alternating vertices $5$ and
$6$ corresponds to the macroscopically smooth $(110)$ plane.
The twofold degeneracy (exchanging vertices
$5\leftrightarrow 6$) represents the invariance of the crystal
surface by the removal of the top layer of atoms.
In terms of the original dipoles, this configuration of vertices
has a zero net dipole moment (which trivially translates
to the facet $(110)$ having zero slope, being the reference
plane). Regions of net dipole moment different
from zero (nonzero slopes $p$, $q$) correspond to tilts away from the
reference plane.
To find $f$ (or equivalently $\overline{f}$)
of this system of interacting dipoles attached to the links of a square
grid, we consider a lattice of $N\times M$ sites and impose
periodic boundary conditions on both directions. Define the
row--to--row transfer matrix
\be
T(u)_{\{\underline{\alpha}\},\{\underline{\alpha^{\prime}}\}}=
\sum_{\{\underline{\beta}\}}\prod_{k=1}^{N}
R_{\alpha_{k} \alpha_{k}^{\prime}}^{\beta_{k} \beta_{k+1}}(u)
\ee
where $\{\underline{\a}\} = \a_1 ,\cdots ,\a_{N}$ 
are the arrows on a row of $N$ vertical links (the $\a^{\pr}$'s
are one row above) and $\b_{k}$ ($\b_{k+1}$) is the arrow to the left
(right) of the k-th vertex. $R_{\a_{k}\a_{k}^{\pr}}^
{\b_{k}\b{k+1}}$ is the Boltzmann weight associated to a given 
configuration  $\{\a_{k},\a_{k}^{\pr},\b_{k},\b_{k+1}\}$ of the
k--th vertex (see fig. $2$). It is more conveniently described
in terms of the so--called spectral parameter
$u$ and anisotropy $\gamma$, whose explicit relation to the original parameters reads
\be
\sinh u = e^{-J_1 / k_{\mbox{\tiny B}}T}\sqrt{\cosh^{2}\gamma - 1}
\ee
\be
\cosh\gamma = -\cosh\frac{1}{k_{\mbox{\tiny B}}T}(J_1-J_2) +
\half e^{\frac{1}{k_{\mbox{\tiny B}}T}(J_1 + J_2)}
\ee
and $k_{\mbox{\tiny B}}$ is the Boltzmann constant. We also incorporate
the Boltzmann factors in new definitions for $H$ and $V$, namely
$h=\beta H$ and $v=\beta V$ (see fig. $2$).

The transfer matrix can be diagonalized exactly with the Bethe Ansatz.
Its eigenvalues read \cite{syy}
\ba
\Lambda(u,\gamma,h,v) &=& e^{v (N-2n)}e^{hN}\biggl[
\frac{\sinh (\gamma - u)}{\sinh\gamma}\biggr]^{N}\prod_{j=1}^{n}
\frac{\sinh(\frac{\gamma}{2} + u - \frac{i\alpha_j}{2})}
{\sinh(\frac{\gamma}{2} - u + \frac{i\alpha_j}{2})}\nonumber\\
&&+e^{v (N-2n)}e^{-hN}\biggl[
\frac{\sinh u}{\sinh\gamma}\biggr]^{N}\prod_{j=1}^{n}
\frac{\sinh(\frac{-3\gamma}{2} + u - \frac{i\alpha_j}{2})}
{\sinh(\frac{\gamma}{2} - u + \frac{i\alpha_j}{2})} \label{def3}
\ea
where the $\a_j$ are the roots of the
Bethe Ansatz equations
\be
\biggl[ \frac{\sinh (\frac{\gamma}{2} + \frac{i \alpha_k}{2})}
{\sinh(\frac{\gamma}{2} -
\frac{i \alpha_k}{2})}\biggr]^N = (-1)^{n+1}e^{2hN} \prod_{l=1}^n
\frac{\sinh(\gamma +
\frac{i}{2} (\alpha_k-\alpha_l))}{\sinh(\gamma -\frac{i}{2}
(\alpha_k-\alpha_l))}\;\;\;\;\;\;k=1,2,\cdots,n
\label{eq10}
\ee
$n$ represents the number of arrows in a row
of $N$ vertical links which are reversed with respect to
the reference ferroelectric state $\ket{\uparrow\uparrow\cdots\uparrow}$.

The specific free energy is given by
$f=-\lim_{N,M\ra\inf}\frac{k_{\mbox{\tiny B}}T}{NM}\ln{\bf Z}$
where ${\bf Z} = Tr (T^M)$ is the partition function.
In the thermodynamic limit this expression
is dominated by the largest eigenvalue $\Lambda_0$ of $T$ and
and the free energy reduces to
\bd
f(u,\gamma,h,v)=-\lim_{N\ra\infty}k_{\mbox{\tiny B}}T
\frac{\ln \Lambda_{0}(u,\gamma,h,v)}{N}
\ed
At this point it is interesting to discuss some already
known facts about the physics of the asymmetric six--vertex
model before solving the Bethe Ansatz Equations explicitly.
The phase diagram and the nature of the phase transitions
are well understood when $h=v=0$ (symmetric six-vertex), or when
$h=0$ and $v \neq 0$ \cite{lieb_wu}.

The $T=0$ phase diagram for $(h,v)\neq 0$ is trivial \cite{syy}.
In this case the ground state wanders, as a function of the
external fields, through each one of the
different regions composed exclusively of vertices of type $1$, $2$,
$\cdots$, $5+6$. The
boundaries between each region are sharply defined. In figure $3$
a plot of the free energy both in the $(h,v)$--plane as well as in the
$(\eta_x,\eta_y)$--plane are presented. The crystal facets to which
each region corresponds are accordingly indicated \cite{jaya_saam}.

As the temperature is raised, thermal excitations bring about a
rounding of edges \cite{van_beijeren}.
When $T\neq 0$ ($\gamma$ finite) the free energy
$f(h,v)$ remains constant as  function of the field (`flat phase')
in a region of the $(h,v)$ plane containing $h=v=0$ and bounded by
a curve $\Gamma$. Beyond this, the field is
sufficiently strong 
to destroy the antiferroelectric order of the system, but not strong
enough to impose ferroelectric order, and an incommensurate phase
appears where the polarization (or surface slope), which is zero
in the flat phase, changes continuously with the field.
The spectrum of the transfer matrix
is gapless with finite size corrections typical of the gaussian model
\cite{noh_kim}. The curve $\Gamma$ has been investigated only at a
few points \cite{jaya_saam,lieb_wu} and the nature of the phase
transition was found to be of
a Pokrovskii--Talapov type \cite{pt}. To study this curve in more
detail, one has to solve the Bethe Ansatz Equations in their limiting
form for arbitrary values of $h$, $v$ and determine the free energy.
This is a non trivial problem. A detailed exposition of the methods
used to tackle (\ref{eq10}) can be found in \cite{nolden, silvio2,
giuseppe}.
Here we point out only the main results which are of relevance
to the questions addressed. Since the Boltzmann factor is
inessential to the forecoming discussion,  we drop it out of our
analysis. The vertical polarization $y$ is defined through
\be
y=\lim_{N\rightarrow\infty}\frac{2S^z}{N}=
\lim_{N\rightarrow\infty}\biggl(1 - \frac{2n}{N}\biggr)
\label{polarization}
\ee
Inside the boundary curve $\Gamma$ in the $(h,v)$--plane
the polarization is constant at its $y=0$ value;
the free energy does not depend on the field and is given by
\cite{syy}
\be
f(u,\gamma,h,v) = -2\sum_{n=1}^{\infty}\frac{e^{-2\gamma n}}{n\cosh \gamma n}
\sinh (nu)\sinh n(\gamma -u)
\label{def22}
\ee
The parametric equation $(h(b),v(b))$ of
$\Gamma$ is given by \cite{nolden,silvio2}
\be
\label{h}
h(b) = -\frac{b}{2}-\sum_{n=1}^{\infty}(-)^{n}\frac{\sinh nb}{n\cosh n\gamma}
\ee
and
\be
\label{v} 
v(b) =\frac{\gamma}{2} -\frac{|\gamma -2u -b |}{2} +
\sum_{n=1}^{\infty}\frac{(-)^n}{n}
\frac{\sinh [n(\gamma-|\gamma-2u-b|)]}{\cosh n\gamma}\;\;\;\;\;\;\;
-\gamma\leq b\leq\gamma
\ee
These equations reproduce only half of
the curve $\Gamma$. The other half can be recovered
from the symmetry $f(-h,-v)=f(h,v)$. Figure $4$
depicts $\Gamma$ in the $(h,v)$--plane along with the same plot in the
$(\eta_x,\eta_y)$--plane (the latter being the boundaries of the
$(110)$ facet) obtained numerically for $\cosh\gamma = 21$, $u=0.5$
(non zero, finite temperature). Two points, $(h_c,v_c)$ and
$(-h_c,-v_c)$ are singled out: the reason for this is explained in what
follows.

It is believed \cite{van_beijeren,jaya_saam,nolden} that the free energy
singularity should be governed by a universal exponent $\frac{3}{2}$.
A calculation performed for $\eta_y = 0$ gives \cite{jaya_saam}
\bd
\overline{f}\sim \overline{f_0} + {\mbox{constant}}\times
(\eta_x - \eta_{x}^{crit})^{3/2} + {\cal O}(\delta\eta_{x}^{2})
\ed
From this result, it follows straightforwardly that
\bd
z = z_0  + {\mbox{constant}}\times (x-x_0)^{3/2}
\ed
An earlier result also support this observation, the direction of
approach being however tilted relative to the normal of the surface
\cite{lieb_wu}.
It remains however to be proven that such results apply irrespective
of the direction of approach and for the whole facet boundary.
In order to prove this, one has to perform expansions on the free
energy by means of an extension of previous techniques
so as to allow an arbitrary angle of approach
\cite{silvio2, giuseppe}.
Consider first the points ($h_c,v_c$). Here the equations simplify
significantly and the increments $\delta h$, $\delta v$ are perpendicular
and tangential to the boundary, respectively. One obtains, when
approaching $\Gamma$ from the incommensurate phase
\ba
f(u,\gamma, h_c+\d h, v_c) &=& f(u, \gamma, h_c, v_c)- 2 c_2 \biggl(
\frac{2}{c_1} \d h\biggr)^{3/2} \nn \\
f(u, \gamma, h_c, v_c + \d v) &=& f(u, \gamma, h_c, v_c) -
\frac{2}{c_{2}^{2}} \biggl(\frac{c_1}{6\pi}\biggr)^3 |\d v|^3
\nn
\ea
where $c_1$ and $c_2$ are functions of $\gamma$ and $u$ only.
These results are not related to the unique geometrical character
of the points. They hold over the whole boundary curve and
can be expressed using the parametric equation
of $\Gamma$, $(h(b),v(b))$ given by eqs. (\ref{h}) and (\ref{v}).
We introduce the following notation \cite{giuseppe}
\ba
v_{t}(b) &=& \frac{d}{db} v(b) \ \ \ \ h_{t}(b)=\frac{d}{db}
h(b) \nn\\
v_{1}(b) &=& \frac{d}{db}v_{t}(b) \ \ \ \ h_{1}(b)=\frac{d}{db}h_{t}(b) \\
\Delta  &=& v_{t}h_{1} - h_{t}v_{1}\nn
\ea
and approach the curve $\Gamma$ from the outside along straight lines
$h=h(b)+ \delta h$, $v=v(b) + \delta v$ where $\delta v = k \delta h$ and
$k$ fixes a slope not tangential to $\Gamma$. The free energy singularity
reads in this case \cite{giuseppe}
\be
\delta f (\delta h) = (\frac{h_{t}}{\Delta})^{1/2} \frac{h_{t}}{3\pi}
[2(k-\frac{v_{t}}{h_{t}})\delta h]^{3/2}
\ee
These results clearly show that the exponent $3/2$ is obtained for all
directions except the tangential one, that is when
$\delta v = \frac{v_{t}}{h_{t}} \delta h $. Here the singularity
is
\be
\delta f(\delta h) = \left\{ \begin{array}{l}
c_{+}\delta h^3\ \ \ \mbox{if}\ \ \  \delta h > 0 \\
c_{-}\delta h^3\ \ \ \mbox{if}\ \ \  \delta h < 0
\end{array} \right.
\ee
where $c_{+} \neq c_{-}$ are constants depending on the parameter $b$ of
the curve.

The result of our long calculation is that one has two critical
exponents $\theta_1 = 3/2$ corresponding to the rounding off of a
crystal facet when the facet boundary is approached
in all other directions except the tangential one and
$\theta_2 = 3$ if the facet boundary is approached tangentially.
Experiments on small Pb-crystals \cite{rottman_wortis2}
which looked for $\theta_1$ have found $\theta_1 = 1.60 \pm 0.15$
whereas experiments on large $\mbox{He}^{4}$-crystals \cite{carmi}
have found $\theta_1 = 1.55 \pm 0.06$ confirming the theoretical
predictions. According to A. Babkin \cite{communication_babkin}
and V. Henrich \cite{communication_henrich} measurements to put
in evidence the second critical exponent $\theta_2$ are feasible.
An open problem common to the measurements in both exponents is the
effect of gravity. These effects have been discussed in \cite{Avron_Zia}.

The authors would like to thank V. Rittenberg for his continuous
encouragement and A. Babkin and V. Henrich for sharing their 
expertise on crystal surface experiments. 
We would also like to thank R. Shankar and D. Kim for the
fruitful discussions. SRD would like 
to thank the hospitality of the Physics
Department at Yale and CAPES (Brazil).

\newpage
\begin{flushleft}
{\Large Figure Captions}\\
\vspace{1cm}
Figure 1.  The fcc cell and vertices. On the top left the
$(110)$ plane is depicted (full circles). Atoms in the plane immediately
above are represented by empty circles. Two such planes combined define the
macroscopically smooth $(110)$ crystal surface. Vertices (full lines, top right)
are defined for each group of $4$ atoms (two filled and two empty circles).
In the middle, a particular configuration of $4$
vertices is depicted, a $-$ sign indicating an atomic vacancy.
The exact correspondence between vertices and configurations of the crystal
plane is shown in the bottom figure. Vertices $1$ through $4$
represent tilts  away from the $<110>$ direction, and correspond locally
to planes $(11\overline{1})$, $(111)$, $(100)$ and $(010)$ respectively.
An alternating configuration of vertices $5$ and $6$ correspond to
the $(110)$ surface.\\
\vskip 0.5cm
Figure 2. The vertices and their corresponding Boltzmann
weights. $\beta$ is the usual Boltzmann factor
$(k_{\mbox{\tiny B}}T)^{-1}$.
\vskip 0.5cm
Figure 3. $T=0$ view of the free energy
in the $(h,v)$-- and $(\eta_x,\eta_y)$--plane.
The values of the parameters are $A=\frac{J_1 -J_2}{2}$,
$B=\frac{J_1 +J_2}{2}$, $C = J_1$, $D=J_2$ and $E=\sqrt{2}(J_1 -J_2)$;
$\widetilde{A}$ is the area of a vertex.\\
\vskip 0.5cm
Figure 4. Parametric plot of $\Gamma$ for
$\cosh\gamma=21$ and spectral parameter $u=\half$
in the $(h,v)$-- and $(\eta_x,\eta_y)$--plane respectively.
The latter corresponds to the boundaries of the $(110)$
fcc facet. The points $(h_c,v_c)$
(lower half--plane) and $(-h_c,-v_c$) (upper half-plane)
are singled out.\\
\end{flushleft}

\newpage
{
\begin{figure}[tb]
\setlength{\unitlength}{1mm}
\def\setl{ \setlength\epsfxsize{12.5cm}}
{
        \makebox{
                \setl
                \epsfbox{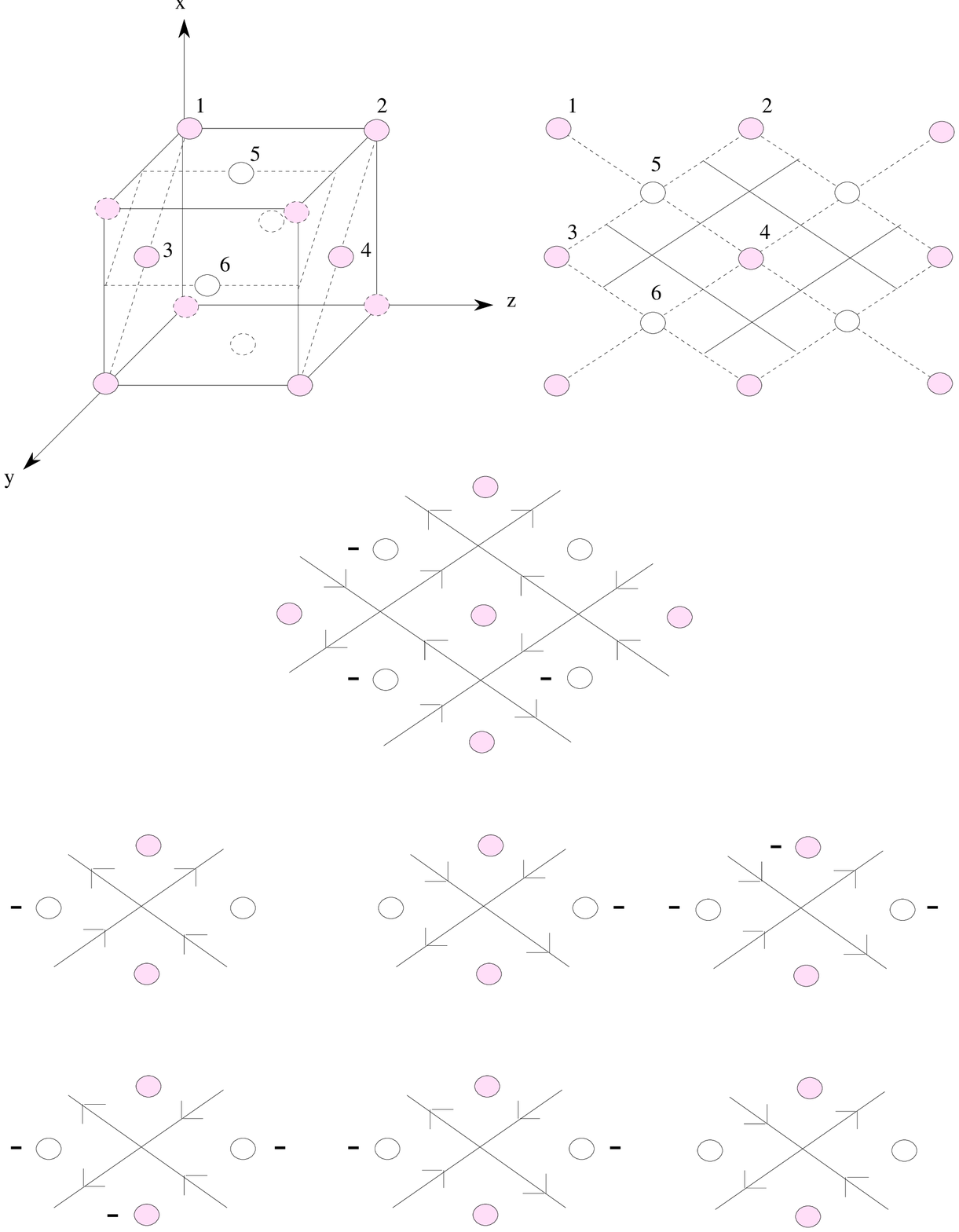}}
        }
\end{figure}
}
\thispagestyle{empty}

\newpage

\pagestyle{empty}
\begin{figure}[b]
\setlength{\unitlength}{5mm}
\begin{picture}(32.5,30)
\put(-1,-1){\framebox(32,7)}
\put(3,2){\vector(0,1){1}}
\put(3,3){\vector(0,1){1}}
\put(2,3){\vector(1,0){1}}
\put(3,3){\vector(1,0){1}}
\put (2.9,1.5) {{\scriptsize 1}}
\put (2.9,4.3) {{\scriptsize 1}}
\put (1.5,2.8) {{\scriptsize 1}}
\put (4.3,2.8) {{\scriptsize 1}}
\put(8,3){\vector(0,-1){1}}
\put(8,4){\vector(0,-1){1}}
\put(8,3){\vector(-1,0){1}}
\put(9,3){\vector(-1,0){1}}
\put (7.9,1.5) {{\scriptsize 2}}
\put (7.9,4.3) {{\scriptsize 2}}
\put (6.5,2.8) {{\scriptsize 2}}
\put (9.3,2.8) {{\scriptsize 2}}
\put(13,3){\vector(0,-1){1}}
\put(13,4){\vector(0,-1){1}}
\put(12,3){\vector(1,0){1}}
\put(13,3){\vector(1,0){1}}
\put (12.9,1.5) {{\scriptsize 2}}
\put (12.9,4.3) {{\scriptsize 2}}
\put (11.5,2.8) {{\scriptsize 1}}
\put (14.3,2.8) {{\scriptsize 1}}
\put(18,2){\vector(0,1){1}}
\put(18,3){\vector(0,1){1}}
\put(18,3){\vector(-1,0){1}}
\put(19,3){\vector(-1,0){1}}
\put (17.9,1.5) {{\scriptsize 1}}
\put (17.9,4.3) {{\scriptsize 1}}
\put (16.5,2.8) {{\scriptsize 2}}
\put (19.3,2.8) {{\scriptsize 2}}
\put(23,3){\vector(0,1){1}}
\put(23,3){\vector(0,-1){1}}
\put(22,3){\vector(1,0){1}}
\put(24,3){\vector(-1,0){1}}
\put (22.9,1.5) {{\scriptsize 2}}
\put (22.9,4.3) {{\scriptsize 1}}
\put (21.5,2.8) {{\scriptsize 1}}
\put (24.3,2.8) {{\scriptsize 2}}
\put(28,4){\vector(0,-1){1}}
\put(28,2){\vector(0,1){1}}
\put(28,3){\vector(-1,0){1}}
\put(28,3){\vector(1,0){1}}
\put (27.9,1.5) {{\scriptsize 1}}
\put (27.9,4.3) {{\scriptsize 2}}
\put (26.5,2.8) {{\scriptsize 2}}
\put (29.3,2.8) {{\scriptsize 1}}
\put (0,0) {$\frac{\sinh(\gamma- u)}{\sinh \gamma}\; e^{h+v}$ }
\put (6,0) {$\frac{\sinh(\gamma- u)}{\sinh \gamma}\; e^{-h-v}$}
\put (12,0) {$\frac{\sinh u}{\sinh \gamma}\; e^{h-v} $}
\put (17,0) {$\frac{\sinh u}{\sinh \gamma}\; e^{-h +v} $}
\put (23,0) {1}
\put (28,0) {1}
\end{picture}
\label{picture}
\end{figure}
\thispagestyle{empty}
\newpage
{
\begin{figure}[tb]
\setlength{\unitlength}{1mm}
\def\setl{ \setlength\epsfxsize{15.5cm}}
{
        \makebox{
                \setl
                \epsfbox{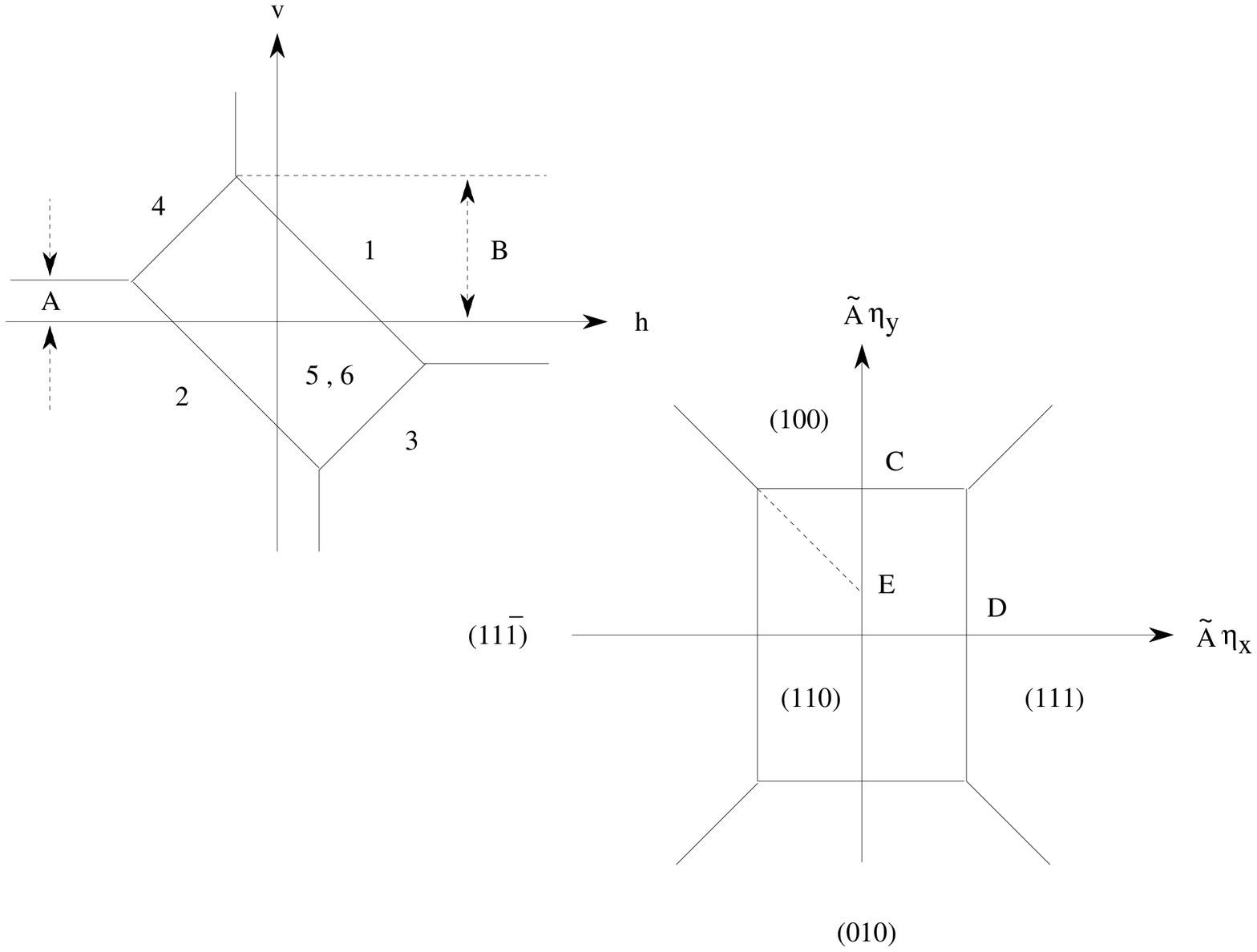}}
        }
\end{figure}
}
\thispagestyle{empty}
\newpage
{
\begin{figure}[tb]
\setlength{\unitlength}{1mm}
\def\setl{ \setlength\epsfxsize{15.5cm}}
{
        \makebox{
                \setl
                \epsfbox{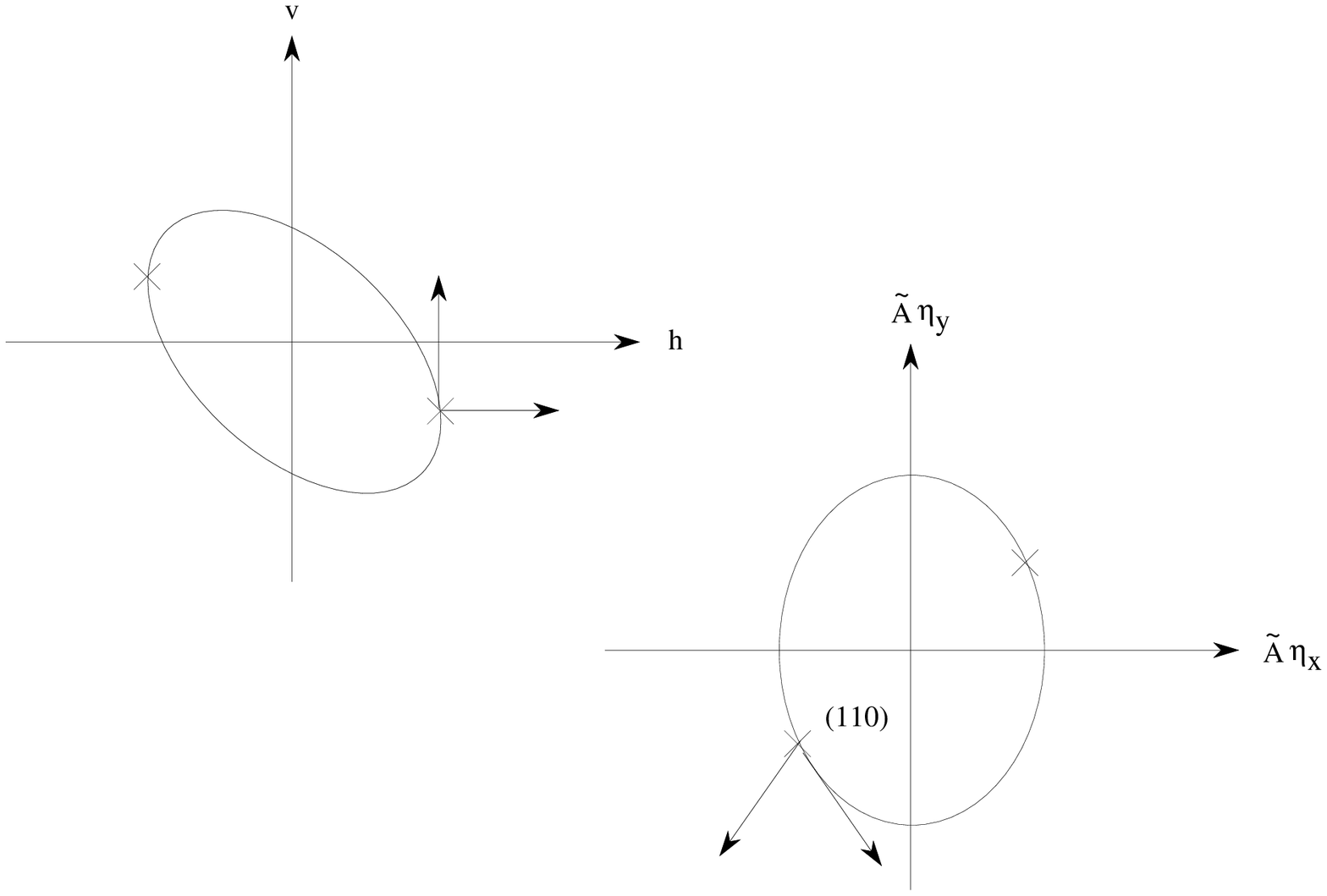}}
        }
\end{figure}
}
\thispagestyle{empty}


\begin{thebibliography}{99}
\bibitem{geometry} The results here presented can be easily
extended to other geometries.
\end{thebibliography}
\end{document}